\documentclass[twocolumn]{aastex6}
 \RequirePackage{lineno}
 \usepackage{amsmath}
 \usepackage{mathrsfs}
 \usepackage{amssymb}
 \usepackage{amsthm}
 \usepackage{natbib,aasdefs,url,bm}
 \usepackage{array}
 \usepackage{float}
 \usepackage{graphicx}
 \usepackage{subfigure}
 \usepackage{color}

\newcommand{\msun}{\rm M_{\odot}}

\begin{document}

\title{Upscattered Cocoon Emission in Short Gamma-ray Bursts \\
as High-energy Gamma-ray Counterparts to Gravitational Waves}

\author{Shigeo S. Kimura\altaffilmark{1,2,3,4,5}, Kohta Murase\altaffilmark{3,4,5,6}, Kunihito Ioka\altaffilmark{6}, Shota Kisaka\altaffilmark{1,2}, Ke Fang\altaffilmark{7,8}, and Peter M\'{e}sz\'{a}ros\altaffilmark{3,4,5}}
\altaffiltext{1}{Frontier Research Institute for Interdisciplinary Sciences, Tohoku University, Sendai 980-8578, Japan}
\altaffiltext{2}{Astronomical Institute, Tohoku University, Sendai 980-8578, Japan}
\altaffiltext{3}{Department of Physics, Pennsylvania State University, University Park, Pennsylvania 16802, USA}
\altaffiltext{4}{Department of Astronomy \& Astrophysics, Pennsylvania State University, University Park, Pennsylvania 16802, USA}
\altaffiltext{5}{Center for Particle and Gravitational Astrophysics, Pennsylvania State University, University Park, Pennsylvania 16802, USA}
\altaffiltext{6}{Center for Gravitational Physics, Yukawa Institute for Theoretical Physics, Kyoto, Kyoto 606-8502, Japan}
\altaffiltext{7}{Kavli Institute for Particle Astrophysics and Cosmology (KIPAC), Stanford University, Stanford, CA 94305, USA}
\altaffiltext{8}{Einstein Fellow}
%
\begin{abstract}
We investigate prolonged engine activities of short gamma-ray bursts (SGRBs), such as extended and/or plateau emissions, as high-energy gamma-ray counterparts to gravitational waves (GWs). Binary neutron-star mergers lead to relativistic jets and merger ejecta with $r$-process nucleosynthesis, which are observed as SGRBs and kilonovae/macronovae, respectively. 
Long-term relativistic jets may be launched by the merger remnant as hinted in X-ray light curves of some SGRBs. The prolonged jets may dissipate their kinetic energy within the radius of the cocoon formed by the jet-ejecta interaction. Then the cocoon supplies seed photons to non-thermal electrons accelerated at the dissipation region, causing high-energy gamma-ray production through the inverse Compton scattering process. 
We numerically calculate high-energy gamma-ray spectra in such a system using a one-zone and steady-state approximation, and show that GeV--TeV gamma-rays are produced with a duration of $10^2-10^5$ seconds. 
They can be detected by {\it Fermi}/LAT or CTA as gamma-ray counterparts to GWs. 
\end{abstract}


\section{Introduction}\label{sec:intro}
The first binary neutron star (BNS) merger event, GW170817, was initially detected by the gravitational waves (GWs) \citep{LIGO17c,LIGO17d}. 
About two seconds later, {\it Fermi}/GBM and {\it INTEGRAL}/SPI-ACS detected the gamma-ray counterpart, which supports the BNS merger paradigm as the progenitor of short gamma-ray bursts (SGRBs)~\citep{LIGO17e}. 
The time lag may imply that the jet launch is delayed for $\sim1$ sec  \citep[e.g.,][]{GNP18a,2019FrPhy..1464402Z}, although other mechanisms can cause time delay~\citep[e.g.,][]{2018PhRvD..97h3013S,IN18a}. The broadband counterparts also confirmed the existence of a relativistic jet \citep[e.g.,][]{MDG18a,MFD18a,2018MNRAS.478L..18T,2019Sci...363..968G,2019ApJ...870L..15L}, which is consistent with an SGRB observed from an off-axis angle \citep[e.g.,][]{IN18a,Troja+18-sgrb150101Bkn, 2019MNRAS.487.4884I}. The UV/optical/IR counterparts \citep{CFK17a,MASTER17a,DES17a,DLT4017a,JGEM17a,SCJ17a,ECK17a,GRAWITA17a,2017ApJ...848L..27T} also verified that a BNS merger produces fast and massive ejecta consisting of {\it r}-process elements \citep[e.g.,][]{KNS17a,KMB17a,MRK17a,SFH17a,JGEM17b}. Gamma-rays above GeV energies and neutrinos are not detected from GW170817 \citep{Fermi18a,HESS17a,IceCube17c}, although they are expected \citep[e.g.,][]{2013PhRvD..88d3010G,KMM17b,2017ApJ...849..153F,MTF18a,KMB18a}.

Light curves of X-ray observations of classical SGRBs have rapidly declining or variable components, such as extended or plateau emissions \citep{NB06a,Swift11a,KBG15a,KYS15a,2019ApJ...877..147K}, which are interpreted as prolonged activities of either a magnetar or a black hole \citep{IKZ05a,PAZ06a,MQT08a,ROM13a,GOW14a,KI15a,KIS17a}. Since duration of the prolonged activities is about $10^2-10^5$ sec, it may be feasible to detect other counterparts of such prolonged emissions with current and near future facilities. However, characteristics of the prolonged jets have yet to be determined.

\begin{figure}
\begin{center}
\includegraphics[width=\linewidth]{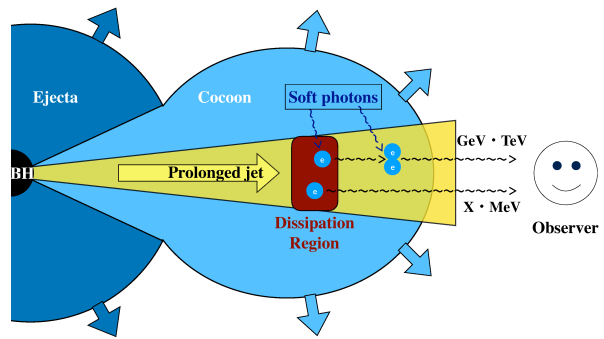}
\caption{Schematic picture of our model. The prolonged jet dissipates its kinetic energy within the cocoon radius. The cocoon supplies soft photons to the dissipation region, leading to GeV--TeV gamma-ray production through the EIC process. Higher-energy gamma-rays are attenuated and reprocessed to lower energies by the cocoon photons before escaping from the system.}
\label{fig:schematic}
\end{center}
\end{figure}

In this Letter, we discuss the upscattered cocoon photons as GeV--TeV counterparts to GWs to probe the prolonged jets. 
Along this line, \citet{MTF18a,2014ApJ...787..168V} proposed using GeV--TeV gamma-ray emissions to probe the prolonged dissipation in SGRB jets, in which interactions between prolonged X-rays and electrons accelerated at external shocks were considered. 
We here consider primary electron acceleration associated with the late dissipation itself~\citep[e.g.,][]{GGN07a,MTY11a}.  
If the ejecta is produced before the jet launch, the jet-ejecta interaction forms a cocoon surrounding the jet \citep[e.g.,][]{Meszaros+01coljet,RCR02a,BNP11a,MI13b,NHS14a,LLC17a,2018MNRAS.473..576G,2019arXiv190905867H}. Since the cocoon freely expands after it breaks out from the ejecta, the dissipation of the prolonged jets may occur inside the cocoon \citep{KIN15a,KIS17a}. 
The cocoon supplies a large amount of photons to the dissipation region, which are upscattered by non-thermal electrons accelerated there (see Figure \ref{fig:schematic}). The upscattered photons interact with the cocoon photons before they escape from the system, which initiates electromagnetic cascades. We calculate the spectrum of GeV--TeV gamma-rays escaping from such a system, and discuss the prospects for future detection as gamma-ray counterparts to GWs. Such an external inverse Compton scattering (EIC) process using the cocoon photons is discussed by \citet{2009ApJ...707.1404T}. They considered the prompt jets with energy dissipation outside the cocoon radius, while we focus on  the prolonged jets with energy dissipation inside the cocoon radius with a more realistic setup. 

We use the notation $Q_X=Q/10^X$ in cgs unit unless otherwise noted and write $Q'$ for the physical quantities in the comoving frame.



\section{The Cocoon}
We estimate the physical quantities of the cocoon in the engine frame. Based on the hydrodynamic simulations of the jet propagation in the ejecta (\citealt{2019arXiv190905867H}; Hamidani et al. in preparation), we set the cocoon mass and average velocity to $M_{\rm coc}\sim 10^{-4}\rm~\msun$ and $\beta_{\rm coc}\sim0.32$. These values are not so sensitive to the jet luminosity. The kinetic energy of the cocoon is estimated to be $\mathcal E_{k,\rm coc}\approx M_{\rm coc}\beta_{\rm coc}^2c^2/2\sim8.9\times10^{48}M_{\rm coc,-4}\beta_{\rm coc,-0.5}^2$ erg, where $M_{\rm coc,-4}=M_{\rm coc}/(10^{-4}\rm~\msun)$.
 
 The thermal energy of the cocoon is initially deposited by the jet-ejecta interaction. Following the simulations by Hamidani et al. (2019, in preparation), we set the initial thermal energy of the cocoon to be a fifth of its kinetic energy: $\mathcal E_{\rm ini}\approx\mathcal{E}_{k,\rm coc}/5\approx1.8\times10^{48}$ erg. For bright prompt jets of SGRBs, the velocity of the jet head is approximated to be $\beta_h\sim1$. Then, the breakout time of the prompt jet is estimated to be $t_{\rm bo}\approx\beta_{\rm ej}t_{\rm lag}/(\beta_h-\beta_{\rm ej})\simeq0.25t_{\rm lag,0}$ \citep{MMR14a,MK18a,2019arXiv190905867H}, where $t_{\rm lag}\sim1$ s is the lag time between the merger and jet launch. After the breakout, the cocoon loses its internal energy by adiabatic expansion. We can obtain the internal energy as $\mathcal E_{\rm ad}\approx\mathcal E_{\rm ini}(R_{\rm bo}/R_{\rm coc})\simeq2.8\times10^{43}\beta_{\rm coc,-0.5}t_{\rm dur,4}^{-1}$ erg, where $R_{\rm bo}\approx t_{\rm bo}\beta_{\rm ej}c\sim1.5\times10^9$ cm is the ejecta radius when the prompt jet breaks out ($\beta_{\rm ej}\approx0.2$ is the ejecta velocity), $R_{\rm coc}\approx t_{\rm dur}\beta_{\rm coc}c\sim9.5\times10^{13}t_{\rm dur,4}\beta_{\rm coc,-0.5}$ cm is the cocoon radius ($t_{\rm dur}$ is the time after the merger). The radioactive decay of {\it r}-process elements also heats up the cocoon. The specific heating rate by the decay chain is expressed by a power-law function for $t_{\rm dur}>1$ s; $\dot\varepsilon_{\rm ra}\approx1.6\times10^{11}t_{\rm dur,4}^{-1.3}\rm~erg~g^{-1}~s^{-1}$ \citep{2012MNRAS.426.1940K,2016MNRAS.459...35H}. The balance between the adiabatic cooling and radioactive heating provides the internal energy to be $\mathcal E_{\rm ra}\approx\dot\varepsilon_{\rm ra}M_{\rm coc}t_{\rm dur}\simeq3.3\times10^{44}t_{\rm dur,4}^{-0.3}M_{\rm coc,-4}$ erg. We write down the cocoon internal energy as $\mathcal E_{\rm coc}\approx\mathcal E_{\rm ad}+\mathcal E_{\rm ra}$. The radioactive heating is the dominant process for $t_{\rm dur}>300$ s with our reference parameter set.

The optical depth of the cocoon is estimated to be $\tau_{\rm coc}\approx 3\kappa_{\rm coc}M_{\rm coc}/(4\pi R_{\rm coc}^2)\simeq53\kappa_{\rm coc,1}M_{\rm coc,-4}\beta_{\rm coc,-0.5}^{-2}t_{\rm dur,4}^{-2}$, where $\kappa_{\rm coc}\sim10\rm~cm^2~g^{-1}$ is the opacity by $r$-process elements. Hence, the photons inside the cocoon should be thermalized. The temperature of the cocoon is written as $a_{\rm rad}T_{\rm coc}^4\approx3\mathcal E_{\rm coc}/(4\pi R_{\rm coc}^3)$, where $a_{\rm rad}$ is the radiation constant. Also, the high optical depth allows us to ignore the photon diffusion effect when estimating the internal energy of the cocoon. Note that the opacity and heating rate in the cocoon may be lower because the neutrino irradiation by the remnant neutron star reduces the amount of lanthanide elements \citep{2018ApJ...860...64F,2018ApJ...856..101M}.

\section{Non-thermal electrons}
\begin{table}
\begin{center}
\caption{Values of the fixed parameters. }\label{tab:fixed}
\begin{tabular}{ccccccccc}
\hline
 $M_{\rm coc}$& $\beta_{\rm coc}$ & $R_{\rm bo}$ & $\epsilon_B$ & $\epsilon_e$ & $p_{\rm inj}$  & $\gamma_m$ \\
 
  [$\rm M_{\odot}$] & & [cm] & & & & &  \\
\hline
  $10^{-4}$ & 0.32 & $1.5\times10^9$ & 0.01 & 0.1 & 2.2 & 100 \\
\hline
\end{tabular}
\end{center} 
\end{table}

\begin{table*}
\begin{center}
\caption{Values of the model parameters and physical quantities. }\label{tab:model}
\begin{tabular}{cccccccccccc}
\hline
 model & $\Gamma_j$ & $L_{k,\rm iso}$ & $R_{\rm dis}$ &  $t_{\rm dur}$ 
 & $\tau_{\rm coc}$ & $\tau_j$ & $R_{\rm coc}$ & $k_BT_{\rm coc}$  & $B'$& $L_{\rm XRT}$ \\
 & & [erg s$^{-1}$] & [cm] & [s] & & & [cm] & [eV]& [G]  & [erg s$^{-1}$] \\
\hline
Extended (EE) & 200 & $10^{50.5}$ & $10^{12}$  & $10^{2.5}$ 
&  $5.3\times10^4$ & 4.6$\times10^{-2}$ & $3.0\times10^{12}$ & 19 & 7.3$\times10^4$& $2.4\times10^{47}$  \\
Plateau (PE) &  100 & $10^{48.5}$ & $10^{13}$  & $10^{4}$ 
& 53 & $3.7\times10^{-4}$ &$9.5\times10^{13}$ & $0.92$ & $1.5\times10^3$& $1.7\times10^{46}$  \\
\hline
\end{tabular}
\end{center} 
\end{table*}

  \begin{figure*}
   \begin{center}
    \includegraphics[width=\linewidth]{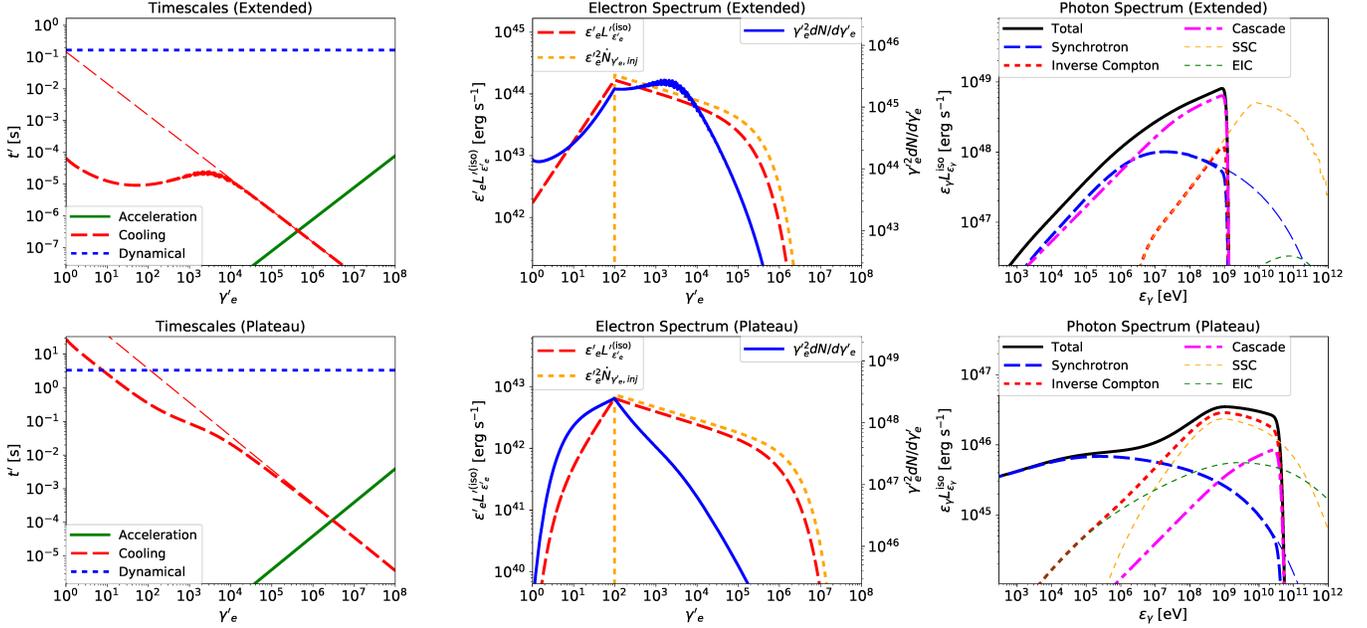}
    \caption{Timescales in the comoving frame (left), electron spectrum at the dissipation region in the comoving frame (middle), and photon spectrum in the engine frame (right) for models of extended (top) and plateau (bottom) emissions. In the left column, the dashed, dotted, and solid lines show the cooling, dynamical, and acceleration timescales. The thin-dashed line is the synchrotron cooling timescale, while the thick-dashed line is the total cooling timescale. In the middle column, the solid, dotted, and dashed lines depict the number spectrum, the injection spectrum, and the differential electron luminosity. In the right column, the thick and thin lines are for the escape and intrinsic photon spectra. We plot the attenuated total (solid-black), attenuated and intrinsic synchrotron (dashed-blue), cascade (magenta-dotted-dashed),  attenuated SSC + EIC (red-dotted), intrinsic EIC (orange-dotted), and intrinsic SSC (green-dotted) spectra. }
    \label{fig:time_dist_lum}
   \end{center}
  \end{figure*}

We consider a prolonged jet with Lorentz factor $\Gamma_j$, isotropic-equivalent kinetic luminosity $L_{k,\rm iso}$, and duration $t_{\rm dur}$. The jet dissipates its kinetic energy at radius $R_{\rm dis}$ through some mechanisms, such as internal shocks \citep{RM94a} or magnetic reconnections \citep{MU12a}. The electron luminosity is set to be $L_{e,\rm iso}=\epsilon_e L_{k,\rm iso}$, leading to the comoving isotropic-equivalent electron luminosity of $L'_{e,\rm iso}\approx L_{e,\rm iso}/\Gamma_j^2$. The comoving magnetic field energy density is given as $U'_B=\epsilon_B L_{k,\rm iso}/(4\pi R_{\rm dis}^2\Gamma_j^2c)$, and the comoving magnetic field is $B'=\sqrt{8\pi U'_B}$. The electron acceleration via diffusive shock acceleration requires the shock upstream region to be optically thin \citep{mi13,KMB18a}. The optical depth is estimated to be $ \tau_j\approx n_j'\sigma_TR_{\rm dis}/\Gamma_j\simeq3.7\times10^{-4}L_{k,\rm iso,48.5}R_{\rm dis,13}^{-1}\Gamma_{j,2}^{-3}$, where $n_j'=L_{k,\rm iso}/(4\pi R_{\rm dis}^2\Gamma_j^2m_pc^3)$ is the comoving number density and $\sigma_T$ is the Thomson cross section. Hence, the electrons can be accelerated in a jet of $L_{k,\rm iso}\lesssim10^{51}\rm~erg~s^{-1}$ and $\Gamma_j\gtrsim100$. The lateral optical depth is estimated to be $\tau_\theta\approx\tau_j\theta_j\Gamma_j\simeq3.7\times10^{-3}L_{k,\rm iso,48.5}R_{\rm dis,13}^{-1}\Gamma_{j,2}^{-2}\theta_{j,-1}$, where $\theta_j$ is the jet opening angle. Thus, the cocoon photons can diffuse into the internal dissipation region as long as the opening angle is small enough.%

The electron distribution in the comoving frame is given by the transport equation that includes injection, cooling, and adiabatic loss terms. Assuming the steady state, the transport equation is written as
\begin{equation}
\frac{d}{d\gamma'_e}\left(-\frac{\gamma'_e}{t'_{\rm cool}}N_{\gamma'_e}\right)=\dot N_{\gamma'_e,\rm inj}-\frac{N_{\gamma'_e}}{t'_{\rm dyn}},
\end{equation}
where $\gamma'_e$ is the electron Lorentz factor, $N_{\gamma'_e}=dN/d\gamma'_e$ is the number spectrum, $t'_{\rm cool}$ is the cooling time, $\dot N_{\gamma'_e,\rm inj}$ is the injection term, and $t'_{\rm dyn}=R_{\rm dis}/(\Gamma_jc)$ is the dynamical time. A solution of this equation is given in Equation (C.11) of \citet{2009herb.book.....D}, and we numerically integrate the solution. Note that the cooling timescale depends on the photon density that is affected by the electron distribution. We iteratively calculate the electron distribution until the solution converges \citep[cf.][]{MTY11a}.

We write the injection term as a power-law function with an exponential cutoff: 
\begin{equation}
 \dot N_{\gamma'_e,\rm inj}=\dot N_{\rm nor}\left(\frac{\gamma'_e}{\gamma'_{e,\rm cut}}\right)^{-p_{\rm inj}}\exp\left(-\frac{\gamma'_e}{\gamma'_{e,\rm cut}}\right),
\end{equation}
where $\dot N_{\rm nor}$ is the normalization factor and $\gamma'_{e,\rm cut}$ is the cutoff energy. The normalization is determined so that $\int \dot N_{\gamma'_e,\rm inj}\gamma'_e m_e c^2 d\gamma'_e = L'_{e,\rm iso}$ is satisfied. The cutoff energy is given by balance between the acceleration and cooling timescales. We estimate the acceleration time to be $t'_{\rm acc}\approx \gamma'_e m_e c/(eB')$. As the cooling processes, we consider the synchrotron, synchrotron-self Compton (SSC), and EIC processes. The synchrotron cooling timescale is estimated to be $t'_{\rm syn}=6\pi m_e c/(\gamma'_e \sigma_T{B'}^2)$. The inverse Compton cooling rate is written in Equations (2.48) and (2.56) in \citet{1970RvMP...42..237B}. We write the differential energy density of seed photons for SSC as 
\begin{equation}
 {U'}^{(\rm ssc)}_{\overline\varepsilon'_\gamma}=\frac{{L'}^{(\rm syn)}_{\overline\varepsilon'_\gamma}}{4\pi R_{\rm dis}^2c},\label{eq:Ussc}
\end{equation}
where $\overline\varepsilon'_\gamma$ is the seed photon energy and ${L'}^{(\rm syn)}_{\overline\varepsilon'_\gamma}$ is the synchrotron differential luminosity (see Section \ref{sec:photons}).
 The seed photons for EIC are the thermal photons in the cocoon boosted by the jet's relativistic motion:
\begin{equation}
 {U'}^{(\rm eic)}_{\overline\varepsilon'_\gamma}=\Gamma_j\frac{8\pi (\overline\varepsilon'_\gamma/\Gamma_j)^3}{h^3c^3}\frac{1}{\exp\left(\frac{\overline\varepsilon'_\gamma/\Gamma_j}{k_BT_{\rm coc}}\right)-1}.\label{eq:Ueic}
\end{equation}
In reality, the photon density and photon temperature in the jet may be slightly lower than those in the cocoon, but we use the photon field in the cocoon for simplicity. This does not strongly affect our results as long as the jet is well collimated and filled with thermal photons.

The left column of Figure \ref{fig:time_dist_lum} shows the cooling and acceleration timescales for models of a typical extended emission (EE) and plateau emission (PE), whose parameters and resulting quantities are tabulated in Tables \ref{tab:fixed} and \ref{tab:model}. 
Note that we here assume a small $R_{\rm dis}$ compared to the previous works \citep[e.g.,][]{2009ApJ...707.1404T,KMM17b} such that $R_{\rm dis}<R_{\rm coc}$ is satisfied. For the EE model, the EIC dominates over the other loss processes for $\gamma'_e\lesssim3\times10^3$, while the synchrotron is the most efficient above it due to strong Klein-Nishina (KN) suppression. For the PE model, the adiabatic loss is dominant for $\gamma'_e\lesssim10$, EIC is efficient for $10<\gamma'_e<6\times10^3$, and synchrotron loss is relevant above it.

The resulting electron spectra are shown in the middle column of Figure \ref{fig:time_dist_lum}. For the EE model, the electron number spectrum, ${\gamma'_e}^2 dN/d\gamma'_e$, shows a hardening for $\gamma'_e\lesssim3\times10^3$ due to the KN effect. For the PE model, the electron spectrum is peaky because of the adiabatic loss below the injection Lorentz factor and the cooling break above it. The cutoff energies are $\gamma'_{e,\rm cut}\sim4\times10^5$ and $3\times10^6$ for the EE and PE models, respectively. We also plot $\varepsilon'_e {L'}_{\varepsilon'_e}^{\rm (iso)}={\gamma'_e}^2m_ec^2(dN/d\gamma'_e){t'}_{\rm cool}^{-1}$ and $\dot N_{\gamma'_e,\rm inj}$ in the figure. The two spectra are almost identical for $\gamma'_e>\gamma'_m$, which confirms the convergence of the numerical integration and iteration.

\section{Gamma-ray spectra}\label{sec:photons}
We numerically calculate photon spectra emitted by the non-thermal electrons. For the synchrotron differential luminosity, we use the formula given by Equations (18), (19), and (20) in \citet{2008ApJ...686..181F}. The spectrum by the inverse Compton scattering is written in Equations (2.48) and (2.61) in \citet{1970RvMP...42..237B}, and the seed photon densities for SSC and EIC are given by Equations (\ref{eq:Ussc}) and (\ref{eq:Ueic}), respectively.

Gamma-rays can be attenuated through $\gamma\gamma$ annihilation inside the system. With our reference parameter sets, the attenuation at the dissipation region is negligible, while the attenuation within the cocoon radius is relevant. Using the quantities in the engine frame, the optical depth for $\gamma\gamma$ interaction is represented by 
\begin{equation}
 \tau_{\gamma\gamma}(\varepsilon_\gamma)\approx\left(R_{\rm coc}-R_{\rm dis}\right)\int\mathscr R(x)\frac{U_{\overline\varepsilon_\gamma}}{\overline\varepsilon_\gamma} d\overline\varepsilon_\gamma,\label{eq:taugamgam}
\end{equation}
where $\varepsilon_\gamma$ is the gamma-ray energy, $\overline\varepsilon_\gamma$ is the soft photon energy in the cocoon,  $x=\varepsilon_\gamma \overline\varepsilon_\gamma/(m_ec^2)$, $\mathscr R(x)$ is given in Equation (4.7) in \citet{cb90}, and $U_{\overline\varepsilon_\gamma}$ is the thermal photon energy density in the cocoon. Note that we focus on the situation where $R_{\rm coc}>R_{\rm dis}$ (see Figure \ref{fig:schematic}). The optical depth above $\sim1$ GeV (100 GeV) for EE (PE) model is very high, so an electromagnetic cascade is initiated. We approximately calculate the cascade spectrum following \citet{1988ApJ...335..786Z}. The cascade spectrum is approximated to be 
\begin{equation}
 \varepsilon_\gamma L_{\varepsilon_\gamma}^{(\rm cas)} \approx \varepsilon_\gamma G(y)
\end{equation}
\begin{equation}
 G(y) = \frac{L_{\rm nor}}{\varepsilon_{\gamma,\rm cut}}\left(y^{3.2}+y^{2}\right)^{-1/4}\exp(-\tau_{\gamma\gamma}),
\end{equation}
 where $L_{\rm nor}$ is the normalization factor, $\varepsilon_{\gamma,\rm cut}$ is the cutoff energy above which $\tau_{\gamma\gamma}>1$ is satisfied, and $y=\varepsilon_\gamma/\varepsilon_{\gamma,\rm cut}$ \citep[see also][]{2012JCAP...08..030M}. $G(y)$ is normalized so that the luminosity of the cascade emission is equal to the energy loss rate by the attenuation.

The calculated photon spectra in the engine frame are shown in the right column of Figure \ref{fig:time_dist_lum}. For the EE model, the injected electrons upscatter the seed photons in the KN regime, so that the EIC process produces high-energy gamma-rays of $\Gamma_j\gamma'_mm_ec^2\simeq10$ GeV. These gamma-rays are absorbed within the cocoon radius and reprocessed as the cascade emission, which is dominant for $\varepsilon_\gamma\gtrsim1$ MeV in the escaping photon spectrum.  The synchrotron emission provides a dominant contribution below it with a hard spectrum. The peak of the synchrotron emission is around 10 MeV due to the strong magnetic field (see Table \ref{tab:model}) and the hard electron distribution. From the observations, the mean value of the photon indices for EEs in the {\it Swift}/XRT band is 1.7 \citep{2019ApJ...877..147K}, which is close to our results of 1.5. For the PE model, the injection energy is in the Thomson regime, and the EIC process produces photons of $\varepsilon_\gamma\gtrsim4\Gamma_j^2{\gamma'_m}^2k_BT_{\rm coc}\sim0.4$ GeV. Above $\varepsilon_\gamma\gtrsim(m_ec^2)^2/(3k_BT_{\rm coc})\sim100$ GeV, the KN suppression is effective. The synchrotron emission produces photons below 20 MeV with a spectral index close to 2, which is also consistent with the observed spectra \citep{2019ApJ...877..147K}. The cascade and SSC emissions are sub-dominant. 
Note that the cutoff energy of the escaping photon spectrum is independent of the Lorentz factor of the jets as long as the electromagnetic cascades in the dissipation region are negligible. Using the cocoon temperature, we have $\varepsilon_{\gamma\gamma,\rm cut}\approx(m_ec^2)^2/(3k_BT_{\rm coc})\simeq4.6$ GeV and 95 GeV for the EE and PE model, respectively. In our cases this overestimates the cutoff energy by a factor of 2--5 because $\tau_{\gamma\gamma}>1$ is satisfied at the exponential tail of the cocoon photons.


\section{Detection Prospects}
We calculate GeV--TeV gamma-ray fluxes from the prolonged jets with various $t_{\rm dur}$ and discuss future detection prospects. The observed EEs and PEs indicate a rough correlation between X-ray luminosity and duration: $L_{\rm XRT}\sim 10^{49}(t_{\rm dur}/10^2~{\rm s})^{2.5}\rm~erg~s^{-1}$ for the EE model and $L_{\rm XRT}\sim 10^{46}(t_{\rm dur}/10^4~{\rm s})^{2}\rm~erg~s^{-1}$ for the PE model, where $L_{\rm XRT}$ is the photon luminosity in $0.3-10$ keV in the observer frame \citep{KIS17a}. By adjusting the value of $L_{k,\rm iso}$, we iteratively calculate the escaping photon spectra for various values of $t_{\rm dur}$ such that the resulting $L_{\rm XRT}$ satisfies the relation above. Here, the XRT band is converted to the engine frame using the mean redshift of the observed EEs, $z=0.72$ \citep{KIS17a}. Our reference models are in rough agreement with the relation (see Table \ref{tab:model}). We focus on 200~s~$<t_{\rm dur}<4\times10^4$~s because $\tau_j>0.3$ and $\tau_{\rm coc}<\beta_{\rm coc}$ are satisfied for $t_{\rm dur}<200$~s and $t_{\rm dur}>4\times10^4$~s, respectively, where our assumptions become inappropriate. Note that the relation between $L_{\rm XRT}$ and $t_{\rm dur}$ strongly depends on the sample and fitting formula \citep{2008MNRAS.391L..79D,2010ApJ...722L.215D,KIS17a,2019ApJ...877..147K}. 

\begin{figure}
\begin{center}
\includegraphics[width=\linewidth]{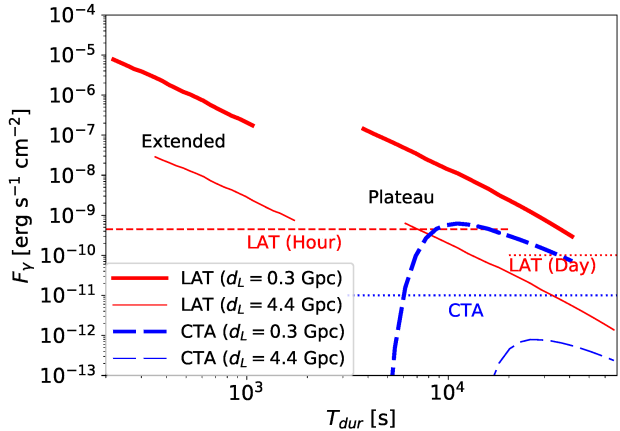}
\caption{Gamma-ray fluxes in LAT (red-solid) and CTA (blue-dashed) bands. The thick and thin lines are for $d_L=0.3$ Gpc and $d_L=4.4$ Gpc, respectively. The red-dashed and red-dotted lines are the LAT upper limit for GW170817 with $\sim10^3$-second integration \citep{Fermi18a} and 1-day integration \citep{MTF18a}, respectively. The blue-dotted line is the sensitivity of CTA for $\varepsilon_\gamma\simeq 50$ GeV for 5-hour integration \citep{2019scta.book.....C}.}
\label{fig:flux}
\end{center}
\end{figure}

In Figure \ref{fig:flux}, we show the gamma-ray fluxes on Earth in the {\it Fermi}/LAT and CTA bands for $d_L=0.3$ Gpc ($z=0.067$; thick line) and $d_L=4.4$ Gpc ($z=0.72$; thin line), which correspond to the GW detection horizon for the advanced LIGO design sensitivity and the average distance to SGRBs with EEs \citep{KIS17a}. The results are depicted as a function of the observed duration, $T_{\rm dur}=(1+z)t_{\rm dur}$. We also plot the CTA sensitivity at 50 GeV \citep{2019scta.book.....C} and the upper limit for GW170817 by LAT \citep{Fermi18a,MTF18a} that mimics the LAT sensitivity. 

GeV gamma-ray emission expected in the EE model is so bright that LAT can easily detect the signals even for $d_L=4.4$~Gpc. Indeed, LAT detected high-energy gamma-rays from two SGRBs, GRB 160702A and 170127C, about a thousand seconds after the trigger~\citep{Fermi19}, which could be explained by our model although detailed fits of individual bursts are beyond the scope of this paper. 
However, LAT has not detected any SGRBs with EEs in $T_{\rm dur}<10^3$ s \citep{Fermi19}, which seems in tension with our model. The cocoon may have a faster and hotter component in its outer region, depending on the initial configuration \citep{2014MNRAS.437L...6K,KNS17a}. The hot component will attenuate the high-energy gamma-rays. Also, if the dissipation region is outside the cocoon radius, which is likely to occur at early times, the EIC flux is significantly reduced. 
These effects need to be considered to explain the non-detection by LAT, and a detailed study remains as a future work. 

In contrast, the PE model is too faint to be detected by both CTA and LAT for $d_L=4.4$ Gpc except for $T_{\rm dur}\lesssim7\times10^3$ s by LAT. For $d_L=0.3$ Gpc, LAT can detect the PEs for the range of durations we investigate here. CTA can also detect the PEs of $T_{\rm dur}<6\times10^3$ s, although it cannot detect the PEs of shorter duration because of its high-energy threshold. Although we focus only on the case with $\tau_{\rm coc}>\beta_{\rm coc}$, the EIC emission may last longer, because the bulk of the merger ejecta (kilonova/macronova) continues to provide seed photons to the dissipation region. 

\section{Summary \& Discussion}
We have considered high-energy gamma-ray emission from prolonged engine activities in SGRBs. We assume that the prolonged jets dissipate their kinetic energy inside the cocoon radius, which provides non-thermal electrons in the dissipation region. The jet-ejecta interaction also produces copious thermal photons, leading to high-energy gamma-ray emission through the EIC process. The calculated photon spectrum is consistent with the X-ray observation, and LAT and CTA can detect gamma-ray counterparts to GWs for duration of $10^2-10^5$ s.
Note that the counterparts by the prolonged engine activities may not be accompanied by the prompt gamma-rays as discussed in \citet{2019Natur.568..198X,MK18a}. Also, Fermi and/or CTA may be able to detect the up-scattered cocoon photons from the SGRBs that occur beyond the GW detection horizon. Hence, the follow-up observations should be performed for GWs without SGRBs and for SGRBs without GWs.

The afterglow and prompt gamma-ray emissions of GW170817 suggest that the jet is structured in which a fast core is surrounded by a slower wing \citep[e.g.][]{LPM18a,2018NatAs...2..751L,MAX18a,2018ApJ...863...58X,2019MNRAS.487.4884I}. If the prolonged jet is structured, the kinetic luminosity at the edge needs to decrease faster than $L_{k,\rm iso}\propto\Gamma_j^{-2}$ in order to fill the jet core with the cocoon photons. The slower edge emits photons to a wider angle, which are reflected by the cocoon. Such photons will be detectable as a wide-angle counterpart to GWs \citep{KIN15a,2018ApJ...867...39K}. Note that the high-energy gamma-rays are not reflected but absorbed by the cocoon through the Bethe-Heitler pair-production process. Thus, LAT or Cherenkov Telescopes would have difficulty detecting the gamma-rays from off-axis events, including GW170817.

The prolonged jets may have a lower Lorentz factor (Matsumoto et al. 2019, in preparation; \citealt{2019ApJ...883...48L}). For the EE model with $\Gamma_j\lesssim 40$, $\tau_j>1$ is satisfied, so the non-thermal particle acceleration does not occur. This condition can be avoided in the jet edge where a kinetic luminosity can be lower, although the emission from the jet edge cannot achieve the observed X-ray luminosity for EEs.
For the EE model with $\Gamma_j\lesssim100$, the GeV gamma-rays are attenuated in the dissipation region due to a higher photon density. 
In this case, if protons are accelerated simultaneously, high-energy neutrinos can be efficiently produced owing to the high target photon density, as has been discussed for X-ray flares and EEs of SGRBs \citep[e.g.,][]{MN06a,KMM17b}. Such neutrinos would be detectable in the planned next-generation neutrino detector, IceCube-Gen2 \citep{Gen214a}. The detection of either high-energy gamma-rays or neutrinos will unravel the Lorentz factor and jet composition, or the non-detection will enable us to place a useful constraint on the emission radius.

\acknowledgments
We thank Hamid Hamidani for useful comments. This work is supported by JSPS Research Fellowship (S.S.K.), KAKENHI Nos. 18H01213, 18H01215, 17H06357, 17H06362, 17H06131 (K.I.), 18H01245, 18H01246, 19K14712 (S.K), Fermi GI program 111180 (K.M., S.S.K., and K.F.), NSF Grant No. PHY-1620777 and AST-1908689, and the Alfred P. Sloan Foundation (K.M.), and the Eberly Foundation (P.M.).



\end{document}